\documentclass{article}
\usepackage{graphics}
 \usepackage{graphicx}

 \usepackage{epsfig}

\usepackage{amssymb}
\usepackage{amsmath}
 \usepackage{amsthm}

\begin{document}

\title{Influence of spin-flip on the performance of the spin-diode}

\author{M.˜ Bagheri Tagani,
        H.˜ Rahimpour Soleimani,\\
        \small{Department of physics, University of Guilan, P.O.Box 41335-1914, Rasht, Iran}}

\maketitle

\begin{abstract}
We study spin-dependent transport through a spin diode in the
presence of spin-flip by means of reduced density matrix
approach. The current polarization and the spin accumulation are
computed and influence of spin-flip on the current polarization
is also analyzed. Analytical relations for the current
polarization and
  the spin accumulation are obtained as a function of polarization of
  ferromagnetic lead and the spin-flip rate. It is observed that the current polarization becomes zero
 under fast spin-flip and the spin accumulation decreases up to $\%85$ when the
  time of spin-flip is equal to the tunneling time.
 It is also observed that the current polarization
  increases linearly when the dot is singly occupied, whereas its
  behavior is more complicated when the dot is doubly occupied.
\end{abstract}

\section{Introduction}
\label{Introduction} Spin dependent transport through mesoscopic
systems has been widely studied both experimentally and
theoretically points of
view~\cite{Mikkelsen,Trauzettel,Lu,Rudzinski,Souza1}. Study of
transport through quantum dots (QDs) provides worthful
information about novel physical phenomena such as Kondo
effect~\cite{Liang,Weymann1}, spin and Coulomb blockade
effects~\cite{Shaji,Inarrea,Park,Dong}, tunneling
magnetoresistance~\cite{Konig,Perfetto,Rudzinski1}, etc. Coupling
of the QD to different leads can result in different transport
properties. Spin filters composed of normal metallic leads
\cite{Engel,Cota} and ferromagnetic leads have been applied for
producing spin accumulation affecting on negative differential
resistance (NDR)~\cite{Weymann2} and zero bias
anomaly~\cite{Weymann3}.
\par An interesting structure is composed of the QD coupled to a
 nonmagnetic lead and a ferromagnetic lead. It has been shown
 that the QD can act as both a spin diode and a spin filter by
 reversing bias voltage~\cite{Rudzinski1,Weymann2}. Diode-like
 behavior in a carbon nanotube coupled to a normal metal and a
 ferromagnetic lead was recently reported ~\cite{Merchant}.
 This behavior was also observed in resonance tunneling connected
 to ferromagnetic contacts with different spin-dependent
 transparencies~\cite{Iovan}.
Spin diodes have been theoretically investigated without taking
into account spin-flip~\cite{Souza,Weymann4}. In this paper, we
analyze this structure with taking into account spin-flip process
in sequential tunneling regime. Rate equations are used to
describe the states inside the QD with considering coherency
between different states. This subject was recently studied by
C.~Feng and co-workers ~\cite{Chi Feng}. Here, we extract
analytical relations for spin current polarization and spin
accumulation. We show how spin-flip decreases the current
polarization and its effect on the spin accumulation is also
studied. Dependence of the current polarization on the
polarization of the magnetic lead is also examined.
\par The article is organized as follows: in the next section,
we compute the spin-dependent current using rate equations. Then,
the numerical results are presented and the behavior of the
system is analytically studied. Finally, some sentences are given
as a conclusion.

\section{Description of the model}
 \label{Model}
We consider a single level quantum dot coupled to a normal metal
and a ferromagnetic lead. Hamiltonian describing the system is

 \begin{equation}
H=H_{lead}+H_D+H_{s.f}+H_T
\end{equation}
where $H_{leads}=\sum_{\alpha k\sigma}\varepsilon_{\alpha
k\sigma} c^{\dag}_{\alpha k\sigma}c_{\alpha k\sigma}$ describes
the leads, $c_{\alpha k\sigma}(c^{\dag}_{\alpha k\sigma})$
destroys (creates) an electron with wave vector $k$, spin
$\sigma$ in the lead $\alpha$ ($\alpha=L, R$).  In continue, it is
assumed that the left lead is normal metal while the right one is
ferromagnetic. Hence energy of
 electron is spin-independent in the left lead, whereas for the
 right lead $\varepsilon_{\alpha k\sigma}=\varepsilon_{k\sigma}-(1)^{\delta_{\sigma
 \downarrow}}\Delta$ that $\Delta$ is band spin splitting.
$H_D=\sum_{\sigma}\varepsilon_d
n_{\sigma}+Un_{\uparrow}n_{\downarrow}$ where $U$ is Coulomb
repulsion, and $n_{\sigma}=d^{\dag}_{\sigma}d_{\sigma}$ is
occupation operator. $d_{\sigma}(d^{\dag}_{\sigma})$ denotes the
annihilation (creation) operator in the QD.
$\varepsilon_d=\varepsilon_0-\alpha_G V_G-\alpha_L V_L-\alpha_R
V_R$ is the energy level of the QD. We assume that the QD is
capacitively coupled to the gate (G), left and right electrodes.
$H_T=\sum_{\alpha k\sigma}[V_{\alpha k\sigma}c^{\dag}_{\alpha
k\sigma}d_{\sigma}+H.C]$ describes the tunneling between leads and
the QD. $H_{s.f}$ contains spin-flip scattering inside the QD,
$H_{s.f}=R[d^{\dag}_{\uparrow}d_{\downarrow}+d^{\dag}_{\downarrow}d_{\uparrow}]$
where $R$ is spin-flip rate. This process can be obtained by, for
instance, spin-orbit interaction~\cite{Danon} or a transverse
magnetic field~\cite{Engel} that rotates the electron spin.
\par The QD can be in i) empty state $|0>$, ii) singly occupied
state $|\sigma>$ ($\sigma=\uparrow,\downarrow$)
($\varepsilon_{|\sigma>}=\varepsilon_d$) or iii) doubly occupied
state $|2>$ ($\varepsilon_{|2>}=2\varepsilon_d+U$). It is clear
that the states $|\uparrow>$ and $|\downarrow>$ are not the
eigenstates of the isolated QD because of spin-flip. In order to
study the QD, reduced density matrix has been used. Using Markov
approximation, time evolution of reduced density matrix elements
are given as follows~\cite{Blum,Mahler} ($\hbar=1$):
\begin{align}\label{Eq.2}
 \frac{dP_{ss'}}{dt}&=-i<s|[H_{s.f},P]|s'>+\delta_{ss'}\sum_{k\neq
 s}(W_{ks}P_{kk}-W_{sk}P_{ss})\\ \nonumber
 & -\frac{1}{2}[1-\delta_{ss'}][\sum_{k\neq s}W_{sk}+\sum_{k\neq
 s'}W_{s'k}]P_{ss'}
\end{align}
$P_{ss}$ is the probability of being in the state $s$
($s=0,\sigma,2$), whereas $P_{ss'}$ describes the coherency
between the states $|s>$ and $|s'>$. $[A,B]=AB-BA$ and $W_{ks}$
denotes the transition from $|k>$ to $|s>$ and is computed by
means of Fermi's golden rule as
\begin{equation}\label{Eq.3}
  W_{ks}=\sum_{\alpha}\Gamma^{\sigma}_{\alpha}[f_{\alpha}(|\varepsilon_{ks}|)\delta_{N_k,N_s+1}+f_{\alpha}^{-}(|\varepsilon_{ks}|)\delta_{N_k,N_s-1}]
\end{equation}
where $N_k$ stands for the number of electrons in the state $|k>$
($N_k=0,1,2$) and $\varepsilon_{ks}=\varepsilon_k-\varepsilon_s$
is the transition energy.
$f_{\alpha}(x)=[1+exp((x-\mu_{\alpha})/kT)]^{-1}$ is Fermi
distribution function of the $\alpha^{th}$ lead with chemical
potential $\alpha$, whereas $f_{\alpha}^{-}=1-f_{\alpha}$. For
computing the tunneling rates ($W_{ks}$), wide band approximation
is used i,.e. dot-lead coupling is energy independent and
$\Gamma_{\alpha}^{\sigma}=\sum_{k}|V_{\alpha k\sigma}|^2$. Hence,
for instance, $P_{00}$ and $P_{\uparrow\downarrow}$ are obtained
from eq.2 as
\begin{subequations}
\begin{align}
\frac{dP_{00}}{dt}&=\sum_{\sigma}[W_{\sigma
0}P_{\sigma\sigma}-W_{0\sigma}P_{00}]\\
\frac{dP_{\uparrow\downarrow}}{dt}&=-iR[P_{\downarrow\downarrow}-P_{\uparrow\uparrow}]-\frac{1}{2}[\sum_{k\neq
\uparrow}W_{\uparrow k}+\sum_{k\neq \downarrow}W_{\downarrow k
}]P_{\uparrow\downarrow}
\end{align}
\end{subequations}
and it is clear that
$P_{\downarrow\uparrow}=P^{*}_{\uparrow\downarrow}$.
\par Solving Eqs. 2 in the steady state ($dP/dt=0$),
spin-dependent current crossing from the lead $\alpha$ is
computed by
\begin{equation}\label{Eq.4}
  I_{\alpha}^{\sigma}=\Gamma_{\alpha}^{\sigma}[f_{\alpha}(\varepsilon_d)P_{00}-f_{\alpha}^{-}P_{\sigma\sigma}
  +f_{\alpha}(\varepsilon_{2\bar{\sigma}})P_{\bar{\sigma}\bar{\sigma}}-f^{-}_{\alpha}(\varepsilon_{2\bar{\sigma}})P_{22}]
\end{equation}
where $\bar{\sigma}$ is the opposite of $\sigma$. For simulation
purpose we set $\Gamma_{L}^{\sigma}=\Gamma_{0}$ and
$\Gamma^{\sigma}_{R}=(1+(-1)^{\delta_{\downarrow,\sigma}}\rho
)\Gamma_0$ where $\Gamma_0$ is the dot-lead coupling strength and
$p$ is the spin polarization degree of the  right lead. We also
set $\mu_L=0$ and $\mu_R=-V$ where $V$ is applied bias. Therefore,
the left lead acts as an emitter when the bias is positive, while
the right one acts as the emitter in negative bias.

\section{Results and discussions}
Fig. 1a shows the current polarization
$\xi=\frac{I^{\uparrow}-I^{\downarrow}}{I^{\uparrow}+I^{\downarrow}}$
as a function of bias. As we expect, spin-flip reduces the
current polarization because this process tries to destroy the
spin polarity of the system. It is observed that the device can
operate as a spin diode when the QD is singly occupied. In
positive bias when $\mu_R<\varepsilon_d<\mu_L$, the current
polarization becomes zero because the left lead acts as the
emitter. Note that $I^{\sigma}=\Gamma_{L}^{\sigma}\rho_{00}$ and
for the left lead, we have
$\Gamma_{L}^{\uparrow}=\Gamma_{L}^{\downarrow}$ and as a result
$I^{\uparrow}=I^{\downarrow}$. On the other hand, for negative
bias i.e. $\mu_L<\varepsilon_d<\mu_R$, the current polarization
is maximum because of
$\Gamma_R^{\uparrow}>\Gamma_{R}^{\downarrow}$. Indeed, a spin-up
electron enters the QD faster than a spin-down electron from the
right lead. Although the device acts as a spin diode in the
presence of spin-flip, the current polarization reduces
significantly. When the QD is singly occupied, the current
polarization is given as~\cite{ref27}
  \begin{subequations}
  \begin{align}
\xi&=0 \qquad \qquad for \quad V>0  \label{Eq.5a} \\
 \xi&=\frac{\rho}{4\alpha^2+1} \quad for \quad V<0 \label{Eq.5b}
\end{align}
\end{subequations}
where $\alpha=R/\Gamma_0$. It is clear from above equation (Eq.
\eqref{Eq.5b}) that increasing $\alpha$ gives rise to decreasing
$\xi$. Note, for $\alpha=0$, our results are the same as Eqs. 21
in the Ref.~\cite{Souza}. If the time of spin-flip is much shorter
than the time of tunneling i.e. $\alpha\rightarrow\infty$, the
current polarization becomes zero. Indeed, spin-flip destroys the
effect of the ferromagnetic lead on the QD, completely.
\begin{figure}[htb]
\begin{center}
\includegraphics[height=70mm,width=70mm,angle=0]{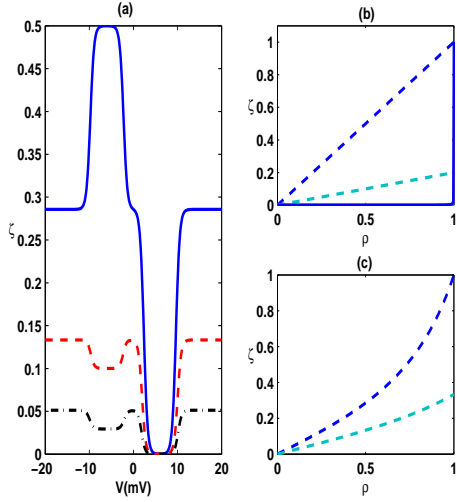}
\caption{(a) The current polarization versus the bias voltage for
$R=0$ (solid), $R=\Gamma_0$ (dashed) and $R=2\Gamma_0$
(dash-dotted). Parameters are $V_G=1mV$, $U=4mV$, $\alpha_G=1$,
$\alpha_L=\alpha_R=0.5$, $T=2.5K$, $\Gamma_0=20\mu eV$ and
$\rho=0.5$.  The current polarization as a function of the right
lead polarization when (b) the QD is singly occupied and (c) the
QD is doubly occupied. $V=6 mV$ (solid) and $V=-6 mV$ (dashed).
We set $R=0$ (blue) and $R=\Gamma_0$ (gray). } \label{fig:1}
\end{center}
\end{figure}
\par The dependence of the current polarization on the $\rho$ under conditions that the QD is singly occupied
 is plotted in the fig. (1b). In positive bias, the current
polarization is always zero, except in $\rho=1$. When the right
lead is a half metal the current polarization will be equal to
one because the spin-down electron injected from the left lead
can not leave the QD, therefore $I^{\downarrow}=0$. This behavior
was reported before in the Ref.~\cite{Souza}. It is
straightforward to show that the probability of being in the
state $|\sigma>$ is given as~\cite{ref27}:
\begin{equation}\label{Eq.6}
P_{\sigma\sigma}=\frac{4\alpha^2+1+(-1)^{\delta_{\sigma,
\uparrow}}\rho}{3[1+4\alpha^2]-\rho^2}
\end{equation}
It is clear from the Eq. \eqref{Eq.6} under conditions that $R=0$,
the QD is occupied by a spin-down electron. In the presence of
spin-flip, this electron can change to a spin-up electron. More
specifically, if spin-flip occurs fast enough ($\alpha\geq 1$), we
have $P_{\uparrow\uparrow}=P_{\downarrow\downarrow}$. As a
consequence, the current polarization is zero under fast
spin-flip even though the right lead is a half metal. In negative
bias, the current polarization increases linearly, as we expect
from Eq. 5b. Increasing $\rho$ means that the spin-up electron
enters the QD faster and ,hence $I_{\uparrow}>I^{\downarrow}$.
The dependence of the current polarization on the $\rho$ when the
QD is doubly occupied is plotted in the fig. 1c. It is observed
that the current polarization is the same for both positive and
negative biases. It can be shown that in high bias
regime~\cite{ref28}, the current polarization is
\begin{equation}\label{Eq.7}
\xi=\frac{\rho}{2[\alpha^2+1]-\rho^2}
\end{equation}
It is observed that when $\alpha=0$ and $\rho=1$, the current
polarization approaches $1$.
\begin{figure}[htb]
\begin{center}
\includegraphics[height=70mm,width=70mm,angle=0]{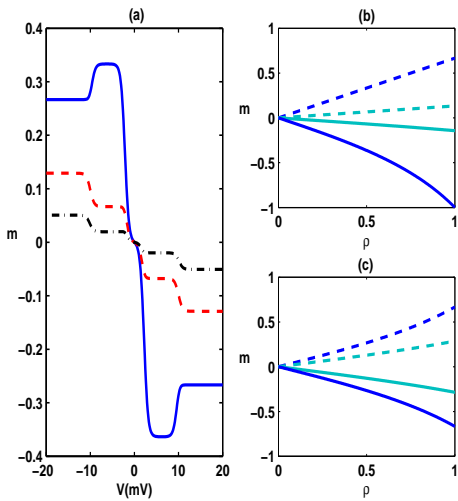}
\caption{(a) The spin accumulation as a function of bias for $R=0$
(solid), $R=\Gamma_0$ (dashed) and $R=2\Gamma_0$ (dash-dotted).
(b) and (c) show the spin accumulation when the QD is  singly
occupied and doubly occupied, respectively. $V=6 mV$ (solid) and
$V=-6 mV$ (dashed).  We set $R=0$ (blue) and $R=\Gamma_0$ (gray).
Other parameters are the same as fig. 1. } \label{fig:2}
\end{center}
\end{figure}
\par The spin accumulation
($m=P_{\uparrow\uparrow}-P_{\downarrow\downarrow}$) is shown in
the fig. 2a. In positive bias, the spin accumulation is negative
because the spin-up electron injected from the left lead leaves
the QD faster and, hence
$P_{\downarrow\downarrow}>P_{\uparrow\uparrow}$. In negative
bias, the right lead operates as the emitter so that the spin-up
electron enters the QD faster and as a result the spin
accumulation becomes positive. Spin-flip scattering reduces the
spin accumulation significantly. In the presence of spin-flip, it
is probable that the electron in the QD  rotates. This rotation
leads to the reduction of the spin accumulation.  There are tow
plateaus in each side of the figure. The first one is due to
entering $\varepsilon_d$ inside the bias window whereas the next
one is created when $2\varepsilon_d+U$ takes place inside the bias
window.
\par The dependence of the spin accumulation on the $\rho$ is
plotted in the  figs. 2a and 2b. It is observed that when the QD
is singly occupied and the right lead is a half metal, the spin
accumulation will be equal to $-1$ if no spin-flip is considered.
This behavior was predictable from Eq. \eqref{Eq.6}. If the
spin-flip rate is equal to $\Gamma_0$, the spin accumulation
decreases up to $\%85$. Unlike positive bias, the spin
accumulation increases linearly in terms of $\rho$ in negative
bias. It can be shown that in single electron state $m$ is given
by
\begin{subequations}\label{Eq.8}
\begin{align}
m&=\frac{2\rho}{\rho^2-3[4\alpha^2+1]} \quad for \quad V>0 \\
m&=\frac{2\rho}{12\alpha^2+3} \qquad \qquad for \quad V<0
\end{align}
\end{subequations}
It is interesting to note that in positive bias, even though it is
assumed that the right lead is a half metal and no spin flip is
considered, $m$ is always lesser than $1$. It results from this
fact that the probability of finding the QD in the empty state is
equal to $1/3$. Indeed, the probability of being in the state
$|\sigma>$ is given as:
\begin{equation}\label{Eq.9}
  P_{\sigma\sigma}=\frac{1+4\alpha^2+(-1)^{\delta_{\sigma,\downarrow}}\rho}{3[1+4\alpha^2]}
\end{equation}
It is clear from above equation that if we set $\rho=1$ and
$\alpha=0$, $P_{\uparrow\uparrow}$ will be 2/3. The magnitude of
$m$ is the same for both positive and negative bias when the QD
is doubly occupied. In this situation, one can obtain~\cite{ref28}
\begin{equation}\label{Eq.10}
  m=\mp \frac{2\rho}{4\alpha^2-\rho^2+4}
\end{equation}
where the minus sign is for positive bias and plus sign is for
negative bias. It is also found that $m$ is always lesser that $1$
because of interplay between spin accumulation and Coulomb
interaction.

\section{Conclusion}
 \label{Conclusion}
In this article, we analyze spin current polarization through a
QD coupled to a normal metal and a ferromagnetic lead by means of
rate equations in the presence of spin-flip scattering. It is
observed when an energy level of the QD is inside the bias window,
this system can operate as a rectifier for spin current
polarization and show how spin-flip can decrease the magnitude of
the polarization. It is observed that the spin accumulation and
the current polarization decrease significantly when spin-flip
occurs faster than the tunneling event. Dependence of the current
polarization and the spin accumulation on the polarization of
magnetic lead is also studied and analytical relations are
obtained as a function of the polarization and spin-flip rate.


\begin{thebibliography}{0}
\bibitem{ref27}
In the singly occupied state, we have $f_{L}(\varepsilon_d)=1,
f_{R}(\varepsilon_d)=0$ in $V>0$, $f_{R}(\varepsilon_d)=1,
f_{L}(\varepsilon_d)=0$ in $V<0$  and
$f_{\alpha}(\varepsilon_d+U)=0$. Hence, $P_{22}=0$ and by solving
the eq.\eqref{Eq.2} and using the normalization relation
($\sum_{s}P_{ss}=1$), Eqs. \eqref{Eq.6} and \eqref{Eq.7} are
obtained.
\end{thebibliography}
\end{document}